\documentstyle[prl,aps,epsf,floats,twocolumn]{revtex}
\begin{document}
\draft
\twocolumn[\hsize\textwidth\columnwidth\hsize\csname @twocolumnfalse\endcsname

\title{Measuring kinetic coefficients by molecular dynamics
simulation of zone melting}
\author{Franck Celestini$^*$ and Jean-Marc Debierre}
\address {Laboratoire Mat\'eriaux et Micro\'electronique de Provence,
Universit\'e d'Aix-Marseille III and CNRS,
Facult\'e des Sciences et Techniques de Saint-J\'er\^ome, Case 151,
13397 Marseille Cedex 20, FRANCE\\
}

\maketitle

\vskip1pc

\begin{abstract}
{ Molecular dynamics simulations are performed to measure the kinetic
coefficient at the solid-liquid interface in pure gold. Results are obtained
for the (111), (100) and (110) orientations. Both Au(100) and Au(110) are in
reasonable agreement with the law proposed for collision-limited growth.
For Au(111), stacking fault domains form, as first reported by
Burke, Broughton and Gilmer [J. Chem. Phys. {\bf 89}, 1030 (1988)].
The consequence on the kinetics of this interface is dramatic:
the measured kinetic coefficient is three times smaller
than that predicted by collision-limited growth. Finally,
crystallization and melting are found to be always asymmetrical
but here again the effect is much more pronounced for the (111) orientation.
}
\end{abstract}


\vskip2pc
]

\section{Introduction}

Solidification of pure elements is of technological interest because
the way a given material solidifies usually affects its structure and,
as a consequence, its final elastic and other macroscopic properties.
   From a fundamental point of view, interest in free and directed solidication
comes from the underlying nonlinear
physics, morphological instabilities being at the origin of generic
microstructures such as dendrites or cells.

Important theoretical and numerical contributions have been made to solve
this difficult physical problem \cite{klaus}. Recently,
a quantitative phase field model was introduced \cite{karma1}. A
subsequent refinement,
consisting in solving the diffusion equation with the help
of Brownian walkers, permitted to bridge the wide gap between
the capillary and diffusion lengths, allowing direct comparison with
experiments \cite{plapp}.
As a consequence, there is currently an increasing need for accurate
values of the interfac
response functions that are used as input parameters for realistic
phase field simulations.

In the case of a pure element, the surface tension $\gamma_{\ell mn}$
must be known as a function of the interface orientation $(\ell mn)$.
In addition, the kinetic coefficient $\mu_{\ell mn}(T_i)$ giving the
relation between
the interface velocity and the interface temperature $T_i$, should also be
known for the different orientations.
For a binary alloy, temperature dependence of the solute diffusion
coefficient, $D(T)$,
as well as velocity and orientation dependence of the
segregation coefficient $k_{\ell mn}(V_i)$ are also necessary.

Both $k$ and $\mu$ are hardly accessible in the experiments and
convection effects
often lead to overestimated values of diffusion coefficients.
Different simulation schemes have
thus been proposed as an alternative. Such numerical experiments have
been rendered possible by
the discovery of realistic interatomic potential models, such as, in
the case of
metals, the embedded atom model (EAM) \cite{eam}, the glue model (GM)
\cite{glue} and the effective medium
theory (EMT) \cite{emt}. In the near future, the increase of computer
power should
open the possibility to address the case of more complicated
materials like semi-conductors, molecular
crystals and organic compounds, for which potentials do not simply
reduce to pair
interactions. Very recently, the functions $\gamma_{\ell mn}$ and
$\mu_{\ell mn}(T_i)$ have
been determined and used in phase field simulations of dendritic
growth for pure
nickel \cite{karma2}. The good quantitative agreement found between experiments
and simulations is promising and should stimulate in the near future
the construction of other
material-dedicated phase field models.

New methods for the determination of the functions $\gamma_{\ell mn}$
and $k_{\ell mn}(V_i)$ have been
recently proposed \cite{asta1,prb}. In the present paper, we rather
concentrate on
$\mu_{\ell mn}(T_i)$. The kinetic response of a solid-liquid interface has been
simulated quantitatively in the 80's by Broughton, Gilmer and Jackson (BGJ)
for a Lennard-Jones (LJ) potential and a (100) orientation
\cite{brou1}. These authors showed
that growth is {\em not} diffusion-limited but rather that the
interface velocity
is related to the mean kinetic energy of the atoms. For this
collision-limited growth  regime,
the growth rate should be directely proportional to the distance
between two successive layers $d_{\ell mn}$.
Indeed, since the liquid atoms do not diffuse to choose their
adsorption sites but are almost
instantaneously incorporated into the solid, the larger $d_{\ell mn}$
the more effective and
the faster the advance of the solid-liquid interface should be. The
analytical expression for the growth velocity of a rough solid-liquid
interface reads\cite{brou2}:
\begin{equation}
       V \propto d_{\ell mn} \tilde V\Big[1-\exp \big(-\frac{\Delta
\tilde \mu}{kT}\big)\Big],
\end{equation}
$d_{\ell mn}$ being the interplane spacing,
$\Delta \tilde \mu$ the chemical potential
difference between solid and liquid phases, $T$ the absolute temperature,
$k$ the Boltzmann constant, and $\tilde V$ the thermal velocity. This
law is confirmed by molecular dynamics simulations for
the (100) and (110) orientations: the expected $\sqrt2$ ratio between
the corresponding kinetic coefficients
is well recovered for several metals cristalizing in a face centered cubic (fcc)
structure (Ni, Ag and Au) \cite{asta2,asta3}.
Nevertheless, for these
rough materials, growth of the (111) interface does not obey this
simple law: according
to Eq. (1), the (111) orientation should be much faster, and what is
found is precisely the opposite.
Burke, Broughton, and Gilmer \cite{brou2} attribute this slowing-down to
the growth of competing fcc and hcp domains in the solidfying layer,
followed by the elimination of the defect lines between the two
phases.

Another question associated with solid-liquid interfaces is that
of symmetry between solidification and melting kinetics. Asymmetry
has been already observed in different systems.
It is not really surprising for faceted materials like silicium
where solidification involves nucleation while  melting does not. The
question is more delicate when one considers rough materials with
collision-limited growth.
Indeed, available results are controversial: if asymmetry has been
found for a Na(100)
interface \cite{ray}, it has not been observed for a LJ(100) \cite{tepper1}.
More surprisingly, in the latter case an opposite asymmetry (crystal
growing faster than the melt) can be found, depending on the way the
solid germ is
prepared.

In this paper, we address the above questions concerning the growth of
a rough solid-liquid interface.  We first present our implementation
of a non-equilibrium molecular dynamics scheme for
a zone melting experiment. The second section is devoted to the study
of (100) and
(110) orientations. The special case of (111) growth is examined in
section III and
asymmetry between melting and solidification in section IV.
Finally, a summary of the different results and a discussion are
given in the last
section.

\section {Simulation procedure}

For this study we use the Ercolessi glue potential for Au \cite{glue}. In
this formalism the total potential energy for a system of $N$ atoms
is given by:
\begin{equation}
U=\frac{1}{2}\sum_{i,j=1}^N \Phi(r_{ij}) +\sum_{i=1}^N U(n_i)
\end{equation}
The first term is a classical pair interaction. In the second term,
$n_i$ is the coordination of atom $i$,
\begin{equation}
n_i=\sum_{j=1}^{N} \rho(r_{ij}),
\end{equation}
where $\rho(r_{ij})$ is a function of the interatomic distance $r_{ij}$,
with a cut-off radius of 3.9\AA\ here.
The energy function $U$ is the {\em glue term} associating an extra potential
energy to  atom $i$ as a function of its coordination. This glue potential has
demonstated its efficiency in predicting the
physical properties of gold as well as in describing several
experimentally observed phenomena such
as surface melting and surface reconstructions \cite{thesis}.

\begin{figure}
\centering
\epsfxsize=3.4cm
\leavevmode
\epsffile{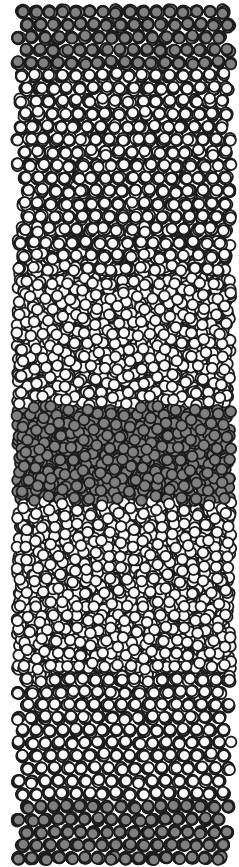}
\vskip1mm
\caption{ A typical simulation box with periodic boundary
conditions in all three directions ((111) solid-liquid interfaces).
Atoms in dark grey are within the
hot and cold slices where
temperature is fixed.}
\label{fig1}
\end{figure}

A distinctive feature of our method is to simulate a zone melting
experiment in which
both a solidification and a melting front are simultaneously
advancing at a fixed velocity $V$.
This velocity is that of the  virtual furnace which imposes two
symmetric thermal gradients.
The particle coordinates are defined in a reference frame
moving at velocity $V$ in the $z$ direction, so that after
equilibration the positions of the two interfaces are fixed in the
simulation box. Heat transport
from the furnace is simulated by imposing one temperature below and one above
the melting point inside two distant slices, 20\AA\ each in thickness (Fig. 1).
Within each slice, temperature is kept constant by
using a classical velocity rescaling procedure \cite{Allen}. Periodic
boundary conditions are applied in  the three directions. More details about
the numerical method can be found in a recent study of solute trapping in
a LJ binary alloy, where a similar simulation technique was used \cite{prb}.

First, the fcc solid and the liquid are equilibrated separatly at zero pressure and
at a temperature close of the melting point. Our smallest system has a size
$S_0\simeq 20\times 20 $\AA$^2$ in cross-section, that is about $64$
atoms per layer.
After equilibration, the solid and liquid are brought into contact
and plunged in the
temperature gradient imposed by the two temperature-controled slices.
The total system is about $220$\AA\ in height.
After a second
equilibration period (during which the velocity of the furnace is
zero), the two
interfaces reach a stationary position and we roughly have $50\%$ of solid
and liquid (see Fig. 1). Fig. 2 shows the temperature and
energy profiles along the $z$ axis.

\begin{figure}
\centering
\epsfxsize=9.0cm
\leavevmode
\epsffile{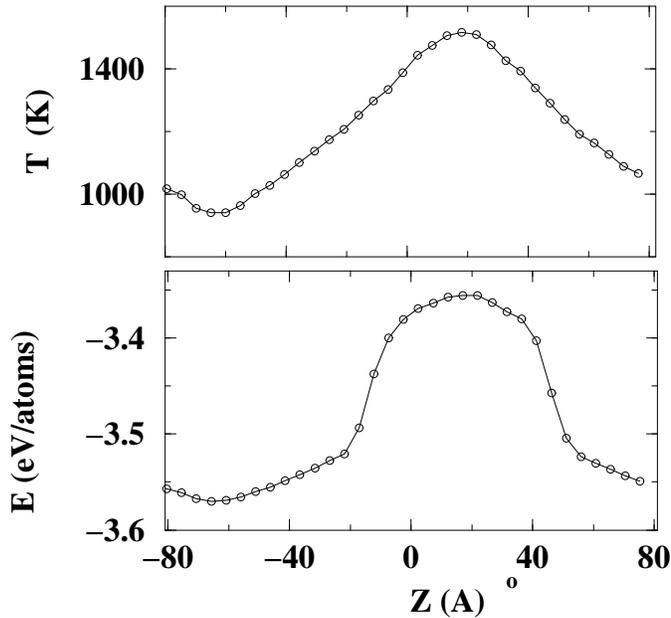}
\vskip1mm
\caption{ Temperature and energy profiles along the $z$ axis
perpendicular to the
interfaces. }
\label{fig2}
\end{figure}

Combining the two profiles to eliminate the $z$ coordinate, one obtains
a caloric curve, {\em i.e.}, a plot of energy as a function of temperature.
In Fig. 3, the caloric curves obtained for two different values of the
pulling velocity V are displayed. For $V=0$, the data points corresponding
to the solidification and the melting fronts merge onto the same
curve: no kinetic
effects are at play and the interface temperature is the equilibrium
melting  temperature $T_0\simeq1330K$. When a velocity is
imposed, a dynamical hysteresis appears on the caloric curve. Kinetic
effects split
the curve in two distinct parts: the interface temperature
of the solidification front decreases while it increases on the
melting front. We can
deduce both interface undercooling and interface superheating from
this plot. An interest
of this method is that, as said before, the interface is fixed in the
reference frame of the
simulation box, so that statistics are easy to record. A tipical run lasts
$10^6$ MD steps ($3.5\times10^3$ ps), so that the atoms in the system
solidify and
melt several times. According to a recent study by Tepper {\em et
al.} \cite{tepper1}, we know that the melting kinetics can be affected
by the way the solid is equilibrated.
Thus multi-cycling is necessary to mimic the melting of a real solid,
usually resulting from previous solidification(s).

\begin{figure}
\centering
\epsfxsize=9.0cm
\leavevmode
\epsffile{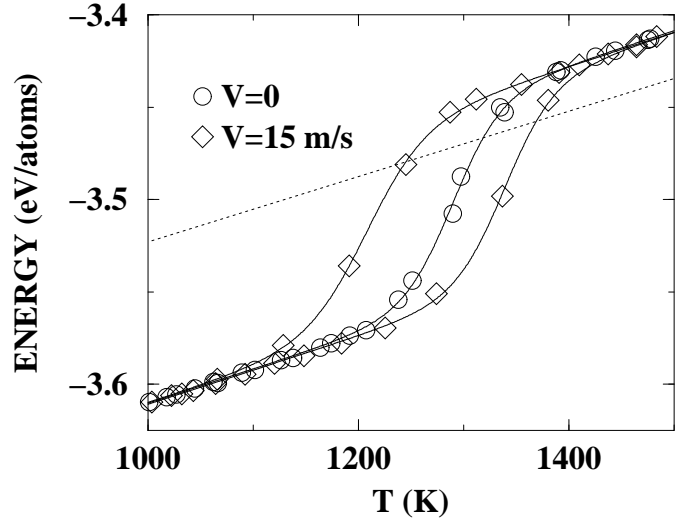}
\vskip1mm
\caption{ Caloric curves for $V=0$ (circles) and $V=15$ ms$^{-1}$ (diamonds).
For non zero velocity,
the kinetic effects split the curve in two parts. The dotted and full
straight lines
represent respectively the functions $E_i(T)$, $E_L(T)$ and $E_S(T)$.}
\label{fig3}
\end{figure}

To conclude this section, the method used to estimate the interface
temperature $T_i$ from the caloric curves is described.
We assume the energy of atoms lying at the interace, $E_i$, to be a weighted
average of the perfect  solid and liquid energies at the same
temperature $T$.
\begin{equation}
E_i(T)=\alpha E_S(T_i)+(1-\alpha)E_L(T_i).
\end{equation}
Linear relations, $E_S(T)=a_ST+b_S$, and $E_L(T)=a_LT+b_L$,
are fitted to the data points obtained on the low and high
temperature side, respectively (Fig.3).
The curve $E_i(T)$ is thus a line with a slope
\begin{equation}
p=\alpha a_S+(1-\alpha)a_L.
\end{equation}
The value of coefficient $\alpha$ is then extracted from the caloric
curve at zero velocity, for which $T_i$ must be equal to $T_0$ (Fig. 3).
 Finally, the interface temperature is given by the intersection of the
 line $E_i(T)$ with the caloric curve. 
An alternative method consists in building an order parameter that
distinguishes between
solid and liquid
atoms \cite{asta1,tepper2}: a plot of this order parameter as a
function of temperature
also gives an interface temperature. We have checked that the two
methods give equivalent
results.

\section{Growth of (100) and (110) interfaces}

\begin{figure}
\centering
\epsfxsize=9.0cm
\leavevmode
\epsffile{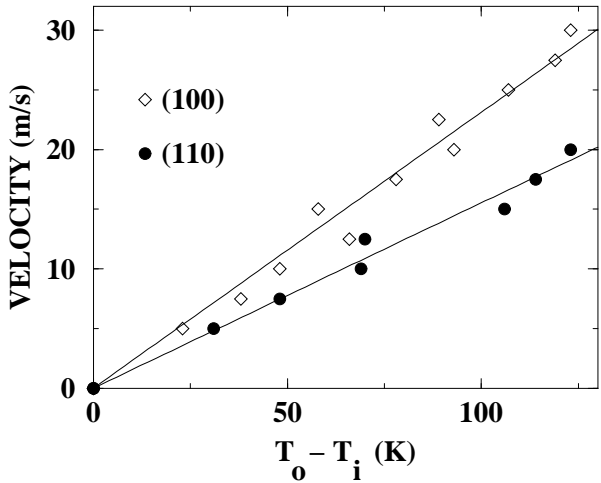}
\vskip1mm
\caption{ Velocity of the solid-liquid interface as a function of
undercooling for
(100) and (110) orientations. The straight lines are best fits to a
linear kinetic law.}
\label{fig4}
\end{figure}

In this section we compute the kinetic coeficient for the Au(100) and
Au(110) interfaces,
using the method described above. We concentrate here on pulling velocities
ranging between $V=5$ $ms^{-1}$ and $V=30$ $ms^{-1}$, for which
kinetics remain linear. We also perform a few simulations at higher
velocities, where kinetics deviates from linearity, but comments on
nonlinear effects are postponed to the concluding section.
In Fig. 4, we plot the interface velocity as a function of the measured
undercooling $T_0-T_i$. Linear fits to the law
\begin{equation}
V=\mu_{\ell mn}(T_0-T_i)
\end{equation}
give the following estimates for the two kinetic coefficients:
\begin{equation}
\mu_{100}^\star=23.1 \pm 1.0 {\rm\hskip2pt cms}^{-1} {\rm K}^{-1}
\end{equation}
and
\begin{equation}
\mu_{110}^\star=15.5 \pm 1.0 {\rm\hskip2pt cms}^{-1} {\rm K}^{-1}.
\end{equation}
However, finite-size effects are expected to bias these estimates
because the system cross-section area, $S_0=20\times20 $\AA$^2$, is
rather small.

Additionnal runs are thus performed in order to quantitatively
evaluate finite-size effects.
The pulling velocity is fixed to $V=15$ms$^{-1}$, the system height
to $H\simeq222$\AA,
and the cross-section area $S$ is progressively increased.
We define the normalized kinetic coefficient $\mu_N(S)$
as the ratio of the  kinetic coefficient obtained at size $S$ to
that obtained at size $S=S_0$ [Eqs (7-8)].
As shown in Fig. 5, the size effects are important and the
kinetic coefficients appear to converge only for $S\simeq100\times100$\AA$^2$.
For the (100) direction, there is a decrease of about $20$ percent and
we obtain roughly the same behavior for the (110) interface. This size effect
has never been reported in the past for (100) and (110) orientations: the fact
that Hoyt and co-workers do no find size effects
for these two orientations \cite{asta1} is certainly due to the fact that their
smaller system is larger than ours. We
can now propose extrapolated values for the kinetic coefficients:
\begin{equation}
\mu_{100}=18.8 \pm 1.0 {\rm\hskip2pt cms}^{-1} {\rm K}^{-1}
\end{equation}
\begin{equation}
\mu_{110}=12.6 \pm 1.0 {\rm\hskip2pt cms}^{-1} {\rm K}^{-1}
\end{equation}
The corresponding ratio
$\mu_{100}/\mu_{110}=1.49\pm0.15$ is in good agreement with the value
$\sqrt2$ predicted
by Eq. (1). Hence, the assumption of collision-limited growth for
(100) and (110) orientations is confirmed to be the relevant one. At
this point,
we can compare our results with those of Hoyt {\em et al.}
for gold \cite{asta3}. If they
also find a $\sqrt2$ ratio between their two orientations, their
$\mu$ values are
larger than ours by a factor $1.8$. Linearizing the expression given by BGJ,
we find
\begin{equation}
V\sim T_0^{-1} T_i^{-1/2} (T_0-T_i)
\end{equation}
for the interface velocity.
The potential used by Hoyt {\em et al.} gives a melting
point $T_0$ of $1090$K \cite{foiles} much smaller than the value
$1330$K obtained with
Eroclessi potential. Introducing this temperature shift in Eq. (11)
roughly accounts for
the discrepancy  between the values of $\mu$. Since Ercolessi potential
gives a melting point much closer to the experimental one, it should
be also the case for our estimates of the kinetic coefficients.

\begin{figure}
\centering
\epsfxsize=9.0cm
\leavevmode
\epsffile{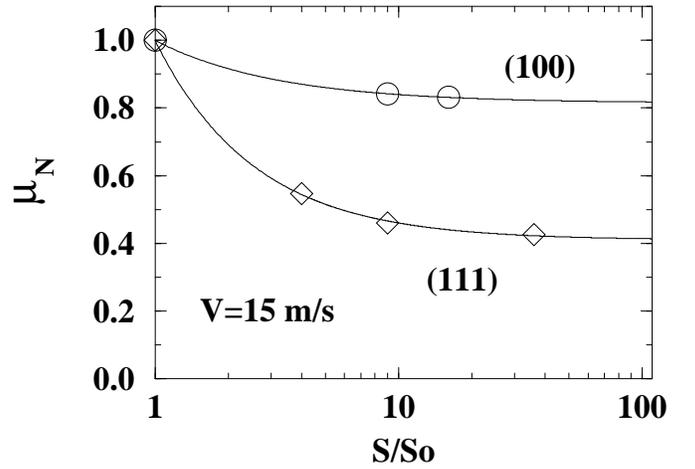}
\vskip1mm
\caption{  Normalized kinetic coefficient as a function of system
size $S$ for the
(100) and (111) orientations.}
\label{fig5}
\end{figure}

In order to understand the origin of the size effects on the value of
the kinetic
coefficient, we take now a closer look at the in-plane structure of
gold layers in the vicinity
of the solid-liquid interface. We compute a density profile along the
$z$ axis from
which we are able to separate atoms belonging to different layers.
Deep in the solid
the in-plane square structure of the (100) orientation is effectively
recovered without
any significant amount of defaults and vacancies. For the two solid
layers just below the
interface the situation is more complex. To distinguish between
different symmetries,
we first perform a Voronoi construction for all the atoms in the layer.
We then collect the set of first neighbors for each atom.

For a fcc solid with lattice parameter $a$, on the square lattice of the (100) orientation
an atom has four nearest neighbors at a distance  $a/\sqrt2$ and 
four second nearest neighbors at a distance $a$.
On the other hand, for a triangular lattice (as the one of
the (111) plane) the six neighbors all
lie at the same distance $a/\sqrt2$. In Fig. 6, we
show a snapshot of the interface solid layer where the Delaunay
triangulation is only drawn for the atoms that have
six first neighbors at comparable distances, in order to reveal
the local triangular structure. It is clear that most of the atoms have
reached their positions on the square lattice but several islands
with a triangular symmetry remain. Note that the number of atoms in the
layer has already attained the value it will have deep in the solid with a
perfect square structure. To compensate for the higher density
of the triangular structure, the corresponding islands are surrounded
by a border
region where the density is very low. This coexistence of two
symmetries is not observed
in our smallest system: one can imagine that for a small area the
square structure is easily
formed and hence triangular islands do not appear. This
phenomenon is very close to the well known reconstruction of the
(100) solid-vapor
interface where the first layer adopts a triangular structure
\cite{daniele}. Turning back
to the solid-liquid interface, the
system apparently uses some of the solidification driving force to
eliminate one of
the two phases and finally reach an almost perfect square symmetry.
Hence, the interface velocity is lower for larger systems.

Such an in-plane ordering is not taken into account in the
collision-limited model but in spite of this we recover the
predicted $\sqrt 2$ value for the ratio $\mu_{100}/\mu_{110}$.
This suggests a similar effect, roughly of the same order, for the (110)
orientation. We have not been able to visualize ordering at (110)
interfaces but one could imagine a mechanism reminiscent of
the missing row reconstruction observed for (110) solid-vapor
interfaces.

\begin{figure}
\centering
\epsfxsize=6.0cm
\leavevmode
\epsffile{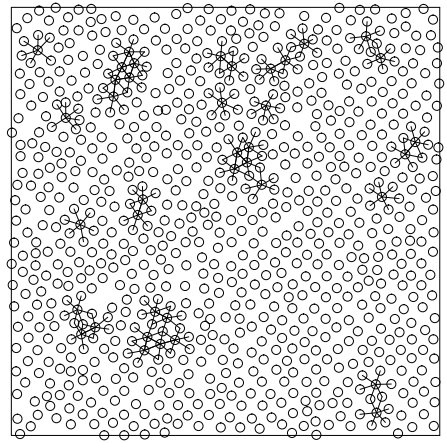}
\vskip1mm
\caption{ Snapshot showing the atoms in the (100) solid layer next to
the interface.
The Delaunay triangulation is only drawn in regions with triangular
underlying symmetry.}
\label{fig6}
\end{figure}

\section{The special case of (111) interface}

We now turn to the case of the (111) orientation. In the same way as
above for the (100) and (110) orientations we calculate the interface
temperature for different velocities. As can be seen in Fig. 7, a
linear kinetic law is also valid for the (111) orientation. Results of
the finite-size analysis, presented in Fig. 5, show that
the size effects are much more pronounced than for the (100) orientation. The
extrapolated value of the kinetic coefficient,
\begin{equation}
\mu_{111}= 7.0 \pm 1.0^{ }  cm s^{-1} K^{-1},
\end{equation}
is now 60 percent below its value for $S=S_0$.
Relatively to the two other orientations, we find
\begin{equation}
\mu_{111} \simeq 0.37 \mu_{100} \simeq 0.56 \mu_{110}.
\end{equation}
These ratios largely differ from the values predicted by Eq. (1), respectively
$2/\sqrt 3\simeq 1.15$ and $2 \sqrt{2/3} \simeq 1.63$. The
(111) orientation,
expected to grow faster because of a larger interlayer spacing, is
surprisingly found to be the slowest one. This discrepency tells us that the
growth mechanism for the (111) orientation is not, or at least not only, a
collision-limited one.

\begin{figure}
\centering
\epsfxsize=7.0cm
\leavevmode
\epsffile{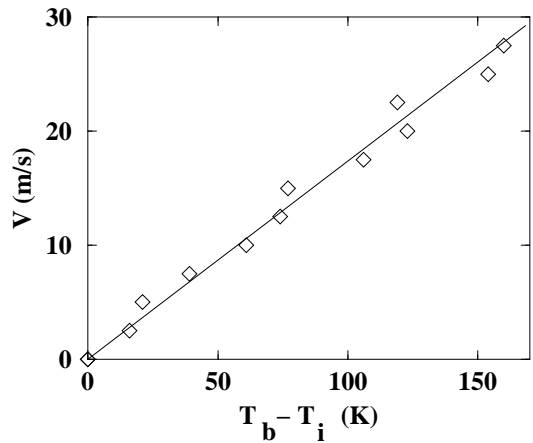}
\vskip1mm
\caption{ Velocity of the (111) solid-liquid interface as a function of
undercooling. }
\label{fig7}
\end{figure}

Here again we look at the symmetries inside the layers close to the
interface. For a (111) layer, there are three possible ordered phases
lying on three different but equivalent sub-lattices that we will call
$a$, $b$ and $c$. As the stacking fault energy is weak for gold
(it is actually zero for the potential we use), once a perfect, say $a$, layer
is formed, the next layer to form is either $b$ or $c$. In Fig. 8, we
show a snapshot
of the three uppermost solid layers and
we distinguish between atoms belonging to $a$, $b$ or $c$ phases. For the
lowest solid layer, phase $a$ is selected and it
occupies the whole plane. For the layer just above, there is
coexistence between $b$ and $c$ sub-lattices. Finally, in the highest
solid layer,
all three phases coexist. We recover here the effect first observed
by Broughton {\em et al.}
\cite{brou2} for a LJ potential. For a (111) orientation the system hesitates
between the different phases it can equivalently form. Here again,
the system dissipates
a part of the available driving force to select one of the phases. As a
consequence, the velocity of the interface is reduced as compared to
the value expected
for a purely collision-limited growth. The size effect is easily
understood because in a small system, coexistence is strongly reduced.
It would be of interest to determine the amount of driving force
spent in this in-plane
organisation in order to estimate the corresponding decrease in $V_{111}$.
To perform this, one could for instance use a 3-state Potts model in three
dimensions  with ferromagnetic intra-plane and anti-ferromagnetic
inter-plane interactions.
To conclude this section we have to point out that phase coexistence is related
to the value of the stacking fault energy $E_s$. For a material with
large $E_s$
phase coexistence should be less probable and the front velocity in better
agreement with the prediction of Eq. (1).

\begin{figure}
\centering
\epsfxsize=6.0cm
\leavevmode
\epsffile{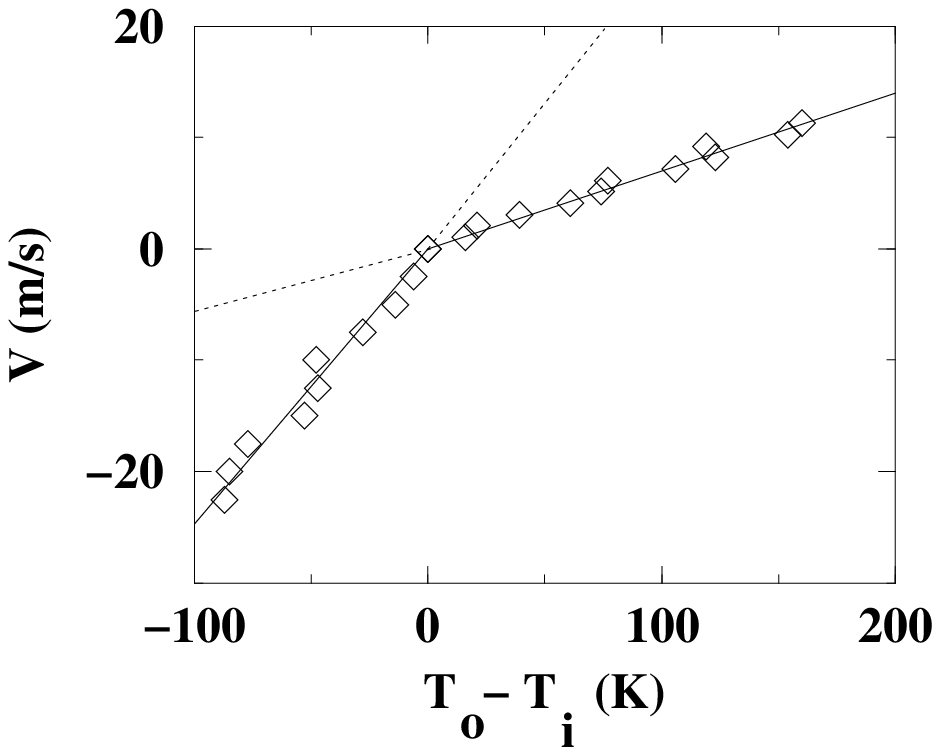}
\vskip1mm
\caption{  Snapshots of the three solid layers immediately
below the inteface (top layer in contact with
the liquid phase). The grey levels correspond to the
three different sub-lattices: $a$ (white), $b$ (light grey), c (dark grey). }
\label{fig8}
\end{figure}

\section{asymmetry between melting and solidification}

As discussed in the introduction, asymmetry is obvious for faceted materials
but is not as clear when considering rough materials like metals.
The question is to know if, at equal absolute undercooling and
superheating, the solid-liquid
and liquid-solid fronts have the same velocity. With our simulation
scheme, this study
is straightforward, since both a melting and a solidification fronts
are simulated at once: no additional calculations are thus required.
Fig. 9 represents
the velocities of both the melting and solidification
fronts as functions of $T_0-T_i$ for the (111) orientation (in our conventions a positive
velocity corresponds
to solidifation). The data are obtained in a system of size $S=S_0$
and corrected according
to the finite-size analysis reported above. It is important to note
that no size effects are
actually found for the melting front: in contrast with the
solidification front,
the melting interface temperature remains the same whatever
the system size. This can be understood if one remembers that for
solidification, especially for
the (111) orientation, growth is not only collision-limited but also requires
in-plane ordering. This is no longer the case for melting, which justifies
the absence of size effects.
The asymmetry shown in Fig. 9 is larger for the (111) orientation.
The same analysis is also made for the two other orientations and we find the
following degrees of asymmetry:
\begin{equation}
\mu_{111}^m=25\pm 4{\rm\hskip2pt cms}^{-1} {\rm K}^{-1} \simeq 3.6 \mu_{111}^s
\end{equation}
\begin{equation}
\mu_{100}^m= 39*\pm 2{\rm\hskip2pt cms}^{-1} {\rm K}^{-1} \simeq 2.1
\mu_{100}^s
\end{equation}
\begin{equation}
\mu_{110}^m =20\pm 2{\rm\hskip2pt cms}^{-1} {\rm K}^{-1}\simeq 1.6 \mu_{110}^s
\end{equation}
where the superscripts $s$ and $m$ refer respectivly to
solidification and melting kinetics.
An asymmetry is revealed in the three cases but it is more
pronounced for
the (111) orientation in the same way as size effects observed during 
solidification.
We conclude here that this asymmetry
is directly related to the ordering within the interface layers. The asymmetry
is strong for (111) because of the peculiar growth mechanism discussed
in the previous section.

The melting front is found to be faster than the solidification
interface in agreement with the
idea that disordering is an easier task than ordering.
Our results confirm the majority of experimental and numerical studies
\cite{ray,kluge,tsao,moss,hunt}. We also confirm the conclusions of
a debate between Richards \cite{richards} and Oxtoby
and co-workers \cite{oxto1,oxto2}  on the importance
of density change on the asymmetry between melting and solidification
kinetics. In agreement with the conclusions of Oxtoby, the gold density change
at melting is small ($\simeq 2\%$)  and can not be responsible for
such an important asymmetry.
On another hand, Tepper
\cite{tepper1} does not find asymmetry for the growth of a (100) LJ solid. Even
if the materials differ, they both belong to the same class of rough
materials and
such a qualitative difference may be surprising. Nevertheless, one
should remember the
strong tendency to surface reconstruction in Au, as observed for the
(100) orientation where
triangular-like regions are formed. This tendency is futhermore
enhanced by the use
of Ercolessi glue potential but is weaker for a LJ potential, what
could explain
the different behaviors observed.

Finally, comparing the melting kinetic coefficients in the different 
orientations,
we find $\mu_{100}^m>\mu_{111}^m>\mu_{110}^m$. We presently do not have
a satisfactory explanation for this hierarchy in the melting kinetics.

\begin{figure}
\centering
\epsfxsize=8.0cm
\leavevmode
\epsffile{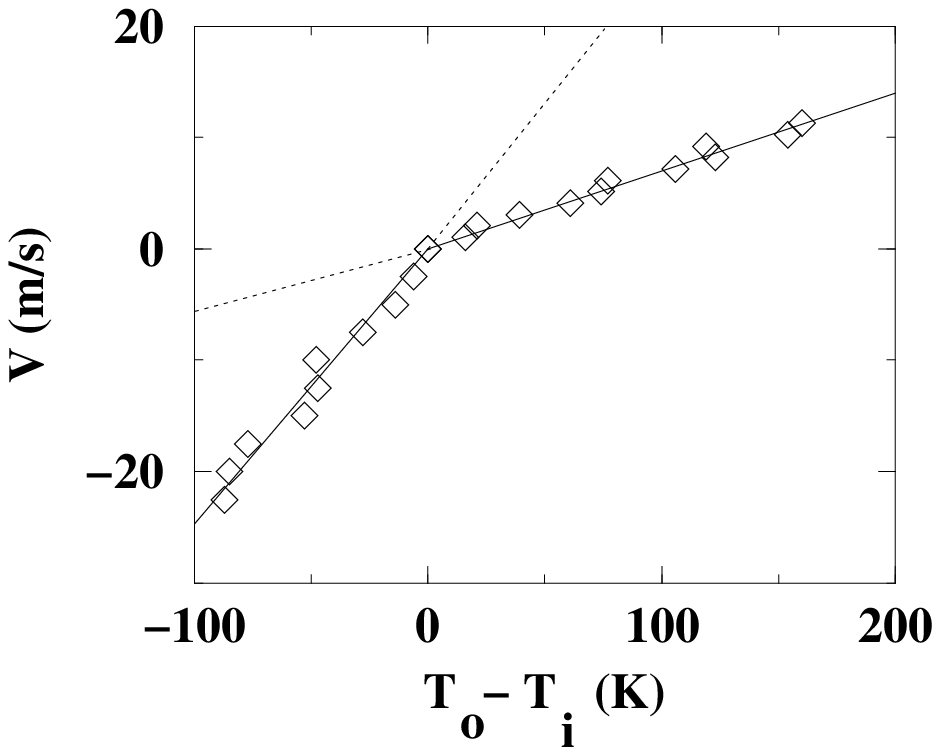}
\vskip1mm
\caption{ Velocity of the (111) solidification and melting interfaces
as a function of
undercooling. Note that the results for solidification incorporate
finite-size corrections.}
\label{fig9}
\end{figure}

\section{Discussion}

Our molecular dynamics simulations of zone melting
experiments allow us to measure simultaneously the solidification and melting
kinetics for a pure element.

For (100) and (110) orientations, growth is apparently well described by a
collision-limited process. Nevertheless, we observe small 2D islands
with triangular
symmetry to form in the solid layer at the (100) solid-liquid interface. As
a consequence,
size effects and asymmetry between melting and solidification are
found. We can not decide
whether this effect is solely due to the tendency of the glue
potential to overestimate surface reconstruction,
or if it is an intrinsic property of gold and/or other metals.
 
The case of the (111) orientation is rather special. Phase coexistence of
three triangular sub-lattices, as first proposed by Broughton {\em et al.}
\cite{brou2}, is recovered. This peculiar behavior has a strong
influence on the kinetics
of the interface. Our finite-size analysis show that in order to
measure a realistic
value of the kinetic coefficient one has to simulate systems with a
solid-liquid
interface area larger than $100\times100$\AA$^2$. The consequence on asymmetry
between melting and solidification is also of importance:
for a given driving force,  the melting front is more than three times faster
than the solidification one.
Because of this disagreement with a purely collision-limited
growth, no analytical model seems, at present, able to predict
the kinetic law of a (111) interface. As discussed previously,
it would be interesting to use a statistical model
to extract the amount of driving force spent for phase separation in order
to modify Eq. (1) and find an acceptable expression for the interface velocity.
 
For melting we find the following order between the different kinetic 
coefficients :
$\mu_{100}^m>\mu_{111}^m>\mu_{110}^m$. To our knoweldge this hierarchy does 
not obey any
existing law. This result will hopefully stimulate further investigations
to reach a clear understanding of the specifities of
melt growth as compared to crystal growth.

The present study is devoted to the linear relationship between velocity and
undercooling. For all the orientations considered here, nonlinear
effects appear
at velocity $V\simeq 30$ $ms^{-1}$ and undercooling $\Delta T \simeq
200 K$. It is not possible
to explain this deviation using either the diffusion limited
\cite{wilson,frenkel} or the collision-limited growth law.
This suggests a possible change in the interface structure
for such large deviations from equilibrium. Density difference
between the liquid and solid phases should also
contribute to trigger  nonlinear behavior \cite{richards}. Understanding
this cross-over would be of importance in the context of very rapid 
solidification.

Finally, we would like to stress that the kinetic effects
can contribute to the anisotropy of the segregation coefficient $k(V)$
for a binary alloy. At sufficiently large velocity, one expects an
important difference
in the interface temperatures for (111) and (100) orientations. As a
consequence, the
diffusivity of solvent atoms and hence the segregation coefficient, 
as predicted
by the Aziz law \cite{Aziz}, should also differ.
This effect may cause solute trapping to appear at lower velocities for
(111) than for (100) or (110) orientations.  We are currently
investigating such
segregation effects induced by kinetic anisotropy in the Al-Cu system.

\acknowledgments

It is a pleasure for us to thank M. Asta, B. Billia,  J. J. Hoyt and A. Karma
for fruitfull discussions.

$*$ Electronic address: celest@l2mp.u-3mrs.fr


\end{document}